# Adsorption of multilamellar tubes with a temperature tunable diameter at the air/water interface

*Anne-Laure Fameau[1,2], Jean-Paul Douliez[2], François Boué[1], Frédéric Ott[1] and Fabrice Cousin[1,*]*

[1] Laboratoire Léon Brillouin, CEA-CNRS, CEA Saclay, 91191 Gif sur Yvette Cedex, France

[2] UR1268, INRA Nantes, Biopolymères Interactions Assemblages, rue de la Géraudière, 44316 Nantes, France

**RECEIVED DATE (to be automatically inserted after your manuscript is accepted if required according to the journal that you are submitting your paper to)**

[*] Corresponding author: fabrice.cousin@cea.fr

**Abstract**

The behavior at the air/water interface of multilamellar tubes made of the ethanolamine salt of the 12-hydroxy stearic acid as a function of the temperature has been investigated using Neutron Reflectivity. Those tubes are known to exhibit a temperature tunable diameter in the bulk. We have observed multilamellar tubes adsorbed at the air/water interface by specular neutron reflectivity. Interestingly, at the interface, the adsorbed tubes exhibit the same behavior than in the bulk upon heating. There is however a peculiar behavior at around 50°C for which the increase of the diameter of the tubes at the interface yields an unfolding of those tubes into a multilamellar layer. Upon further heating, the tubes re-fold and their diameter re-decrease after what they melt as observed in the bulk. All structural transitions at the interface are nevertheless shown to be quasi-completely reversible. This provides to the system a high interest for its interfacial properties because the structure at the air/water interface can be tuned easily by the temperature.

# I Introduction

The main aim of the *green chemistry* is to use renewable biological compounds for application purposes. Thus, there has been recently a growing interest for the industrial valorization of long chains saturated or/and hydroxylated fatty acid. Such hydroxylated fatty acid are indeed very common in nature and can be extracted from agricultural resources, for example from plant biopolymers like cutin [1, 2] and suberin [3] or derivated from castor oil like the 12-hydroxy-stearic acid. Their main drawback for applications comes from their lack of solubility in aqueous media. However, it has been demonstrated recently that with the help of specific counter-ions (choline[4], tetrabutylammonium hydroxide [5], ethanolamine [6], lysine [7] ), the dispersion in water of long chain saturated fatty acids and their hydroxylated derivatives becomes possible, making them good candidates as new green detergents.

For the 12-hydroxy-stearic acid in the presence of ethanolamine as the counter-ion, a very original supramolecular assembly is obtained: fatty acid salts are forming concentric stacked bilayers that are rolled up to form the walls of tubes of a micron size [6]. The system has thus a hierarchical spatial structure, as it is made at local scale of bilayers of fatty acids of ~ 3-4 nm separated by a layer of water of a few tens of nm whereas at larger scale the rolling of the lamellar phases create very long tubes, with a diameter of the order of the 0.6 µm and a length of around 10 µm. Interestingly, if the self assembly of tubes has been already widely observed using various surfactants [8-13], such tubes made of 12 hydroxy stearic acid exhibit a remarkable additional behavior since the outer tube diameter can be increased by a factor ~ 10 in a span of a few degrees. The diameter of the tubes has a value of around 0.6 µm from 20°C to 47°C, a temperature at which it starts to strongly increase to reach a maximum value of 5 µm at 50°C. Then, it decreases again upon a further increase of temperature [14, 15]. Finally, at 70°C, the solution becomes isotropic and no more tubes are longer observed, a disappearance probably linked to the melting of tubes into micelles. At a local scale, the interlamellar spacing also changes with temperature but do not follow the same trends as the diameter. It slightly increases from 35 nm at 20°C to 38 nm at 45°C and drops down to ~ 27 nm at 60°C. This could be related to the thermodynamical ($L_\beta$) gel/ ($L_\alpha$) fluid transition of the fatty chains, measured at 46°C by calorimetry [14]: up to 45°C, the bilayer thickness is about 40 Å, which corresponds to twice the length of the fatty acid chain in its extended conformation, showing that the fatty acids are embedded in a gel bilayer phase ($L_\beta$) while it decreases drastically down to a value of 23 Å above 50°C, a thickness markedly lower, when the fatty acids are in fluid phase ($L_\alpha$).

Besides their temperature properties, these tubes can also survive to important changes of various experimental parameters such as pH and salt [14]. This provides them an evident potential interest in medical applications, for encapsulation and control release [16], or in material science for templating [13, 17]. The solutions of tubes have also enhanced interfacial properties: they exhibit better foaming and emulsifying properties when the fatty chains are assembled in tubes rather than micelles [15].

The understanding of such peculiar interfacial behavior is not trivial and raises the following questions: Are the tubular structures stable at a water/hydrophobic interface (*e g* air in the case of foams)? Or, more generally, what is the relation between the structure of the fatty chains in bulk solution and the structure of the fatty chains at a water/hydrophobic interface? These are the questions we address in this paper by presenting an exhaustive study of the structural behavior of a solution of ethanolamine salt of 12-hydroxy-stearic acid at the air/water interface by coupling surface tension measurements and Neutron Reflectivity (NR). Indeed, it has been demonstrated that NR is a powerful technique for the study of surfactant adsorption for various interfacial architectures such are surfactants [18-24] monolayers, lamellar phases in the vicinity of the interface [25] or multilamellar vesicles [26].

## II Materials and Methods

*Sample preparation*

12-hydroxy stearic acid (Sigma-Aldrich, 99% purity) was weighted accurately in a test tube in which ultra pure water was added to reach a concentration of 10mg/mL (1%). Then, we incorporated to reach equivalence (12-hydroxy stearic acid/counter-ion molar ratio R=1/1), the desired volume of a 1M stock solution of the counter-ion, 2-amino-1-ethanol (Sigma-Aldrich, 99% purity). The mixture was heated at 80°C for 15 min until all components were dispersed and then vigorously vortexed. Prior to use, each sample was heated at 80°C for 15 min and cooled at room temperature.

For the specular neutron reflectivity experiments, the sample was prepared in 3 different mixtures of $H_2O/D_2O$: 40%$H_2O$/60%$D_2O$, 20%$H_2O$/80%$D_2O$ and 100% $D_2O$.

*Specular Neutron Reflectivity*

Specular neutron reflectivity (SNR) experiments were carried out on the horizontal time-of-flight reflectometer EROS at the Laboratoire Léon Brillouin (CEA Saclay, France). The horizontal collimated beam is deflected by a neutron supermirror by an angle of 0.75° on the sample to collect data at a fixed incidence angle of 1.5°. With a neutron white beam covering wavelengths from 3 to 25 Å, the accessible Q-range is 0.01 - 0.1 Å$^{-1}$.

The sample was placed in a sealed cell, with two quartz windows allowing the passage of neutrons, avoiding the exchange of $D_2O$ with $H_2O$ from atmosphere. Measurements were performed at ambient pressure. The temperature of the cell was precisely fixed by a circulating circuit connected with a thermalized Lauda bath. Prior to each measurement, the sample was rest for 30 minutes to reach equilibrium. The acquisition of the SNR data was then recorded for 6 hours with slices of 1 hour.

Most of the experiments were performed during a first series of experiments where the data were acquired by a 1-D detector. It appeared afterwards that there was an intense off-specular scattering in the system that could perturb the determination of the specular scattering. In order to circumvent this difficulty, we performed afterwards some experiments with a 2-D detector to discriminate between specular and diffuse scattering. For lack of beamtime only one sample has been measured with the 2-D detector. The intense diffuse scattering raises the question of the exact determination of the specular scattering because this latter may be contaminated if it is integrated over a too large surface of the detector[27]. This is illustrated in Figure SI.2 that compares 3 different areas of integration of the specular signal of a solution at 10mg/ml in pure $D_2O$, with respective thicknesses of 5 mm, 10 mm and 20 mm (see Supplementary Information). While they show the same characteristic fringes, the reflectivity increases when the integration area increases, and progressively deviates from the Fresnel curve as off-specular signal it integrated in the specular window. It appears that the choice of the integration area does change neither the shape nor the intensity of the fringes. In such a representation, the additional diffuse scattering increases the level of the baseline of the whole curve above the Fresnel reference curve. When the integration window is properly defined the SNR curve follows the Fresnel decay in $Q^{-4}$. We compare also in Figure SI.2 the results obtained with the 1-D detector, which has a thickness of 10 mm. It clearly appears that it slightly integrates too much diffuse scattering (its level is of the order of the 2$^{nd}$ integration on the 2-D detector). This moderate excess of scattering however does not prevent a correct qualitative interpretation of the data obtained with the 1-D detector.

The reflectivity curves corresponding to the analytical models presented within the text are calculated by the optical matrix method with a slicing of the scattering length density profiles in slabs of 10 Å. The experimental resolution of the spectrometer was taken into account in the calculation.

The calculation of the Scattering Length Density (SLD) $N_b$ of 12-hydroxy stearic acid gives -0.28*10$^{-6}$ Å$^{-2}$ with a molar density of 300 g/mol. The respective SLD of the 3 different $H_2O$/$D_2O$ mixtures are 3.61*10$^{-6}$ Å$^{-2}$ for 40%$H_2O$/60%$D_2O$, 4.8*10$^{-6}$ Å$^{-2}$ for 20%$H_2O$/80%$D_2O$ and 6.3 10$^{-6}$ Å$^{-2}$ for 100%$D_2O$.

III.1 **Results and Discussion**

We present here the Specular Neutron Reflectivity (SNR) measurements performed at the air/water interface on solution of fatty acid tubes to determine the corresponding interfacial structure of the fatty acids chains. We have chosen to work at a solution concentration of 10 mg/ml (with a molar ratio of 1/1) because such concentration corresponds to the lower limit for which there are tubes with a temperature-dependent diameter in bulk solution [ref].

Prior to SNR experiments, we have performed dynamic surface tension measurements to determine the characteristic time taken by the fatty chains to reach a stable structure at equilibrium at the air/water interface. The results, shown in Figure SI.1 of the Supporting Information, shows that the system is close to equilibrium after ~ 5 seconds. The evolution of the surface tension in the early 5 seconds suggests that the structure at equilibrium is different from a simple layer of fatty acids monomers at the air/water interface. When we performed the SNR measurements, the solution of tubes was placed in the measurement cell and the samples were let to rest for 30 minutes before starting data acquisition (see Materials and Methods). The system was thus at equilibrium during the experiments (see supporting Information).

*III. 1 Raw data at 20°C*

The curve of Figure 1.a shows the data obtained for a solution in pure $D_2O$, heated at 20°C out of beam and deposited on the preheated cell at the same temperature (see Materials and Methods section). It is compared with the Fresnel curve that corresponds to the pure air/$D_2O$ interface. The same results are also displayed in the Fresnel representation ($R(Q)Q^4$ *versus* $f(Q)$) in Figure 1, which enables to focus on the scattering coming only from the layer present at the interface by compensating the $Q^{-4}$ scattering decay from the pure air/water interface. The displayed curve is the sum of 6 spectra (6 hours). Indeed, 6 consecutive spectra (not shown) have been recorded kinetically by slices of one hour up to a total duration of 6 hours. They were all similar, confirming that the structures adsorbed at the interfaces are stable with time. It immediately appears that there are regular Kiessig fringes Kiessig réguliers pas ? in the specular signal, reflecting that there is a large layer of fatty acid materials at the interface. The raw 2-D spectra reveal that there is also a high diffuse scattering (see Supporting Information). This intense diffuse scattering may come from large spatial heterogeneities on the surface, from the presence of large adsorbed tubes or possibly from the fluctuations of the adsorbed architectures at the interfaces. Please note that no structural ordering at large scale was observed from these off-specular scattering. As explained in the Materials and Methods section, this diffuse scattering does not allow the exact determination of the specular scattering, in particular with the 1-D detector. It nevertheless not hampers the exploitation of all of the data presented in this paper which were obtained with the 1-D detector because we will further show that the useful information is derived from the shape and the positions of the fringes and that the extra diffuse scattering can be taken into account.

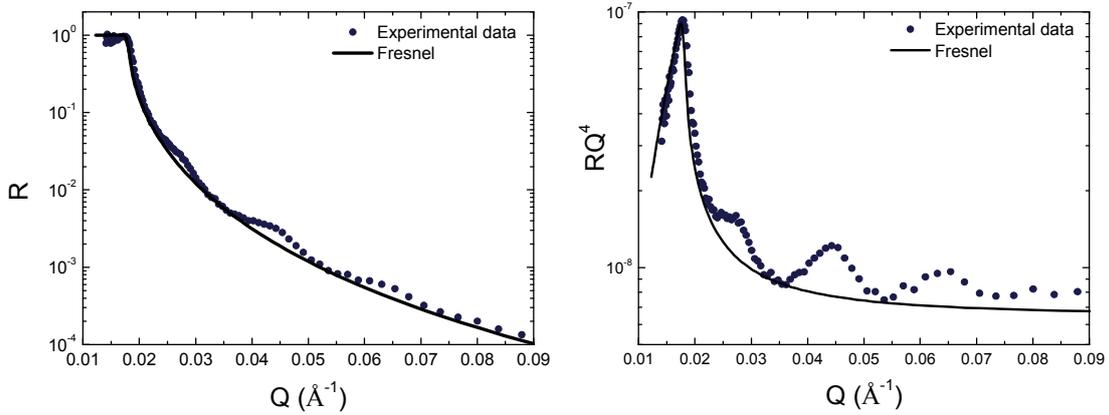

Figure 1: Specular Neutron Reflectivy results of a solution of fatty acids (10 mg/ml) at 20°C at the air/water interface. The solid line is a calculation of the Fresnel reference SNR of a pure air/$D_2O$ interface. (a) R(Q) *versus* f(Q); (b) $Q^4$R(Q) *versus* f(Q).

A calculation of the thickness of the fatty chain layer from the d-spacing of the Kiessig fringes (d ~ $2\pi/\Delta Q$) gives d ~ 300 Å, which is of the order of the interlamellar spacing as measured in the bulk solution (340Å from SANS experiments [6]). Then, the SNR signal may arise from the surface scattering of fatty acid materials and/or from the intense SANS scattering of the tubes in solution below the interface. In order to discriminate between these two possibilities we have performed the same experiments at 20°C with 3 different $H_2O/D_2O$ compositions of the aqueous solvent. The SNR curves are compared in Figure S.I 4. Their shapes are very similar, displaying some fringes that have the same d-spacing and the same amplitude. However, from one sample to another, the fringes are all shifted by a value equal to the critical wave vector $Q_c$ corresponding to the water composition. If the signal was due to bulk scattering, the Q-position of the fringes would have been constant with the water composition, only the amplitude of the fringes would have changed. Then, the difference observed between the data for the 3 $H_2O/D_2O$ contrasts unambiguously prove that the scattering indeed arises from a surface scattering of fatty acid materials. It remains however possible that a small remaining part of the measured scattering comes from SANS.

*III.2 Modeling of the SNR data at 20°C:*

In order to fit our SNR data, we used various models of fatty acid arrangements at the air/water interface. We limit here this modeling to the results of the 2-D detector integrated with the lower surface area (see Materials and Methods), for which the effects of the off-specular scattering do not perturb too much the SNR data. First, since the surface scattering comes from fatty acid materials adsorbed at the air/water interface, the simplest model to be tested is the presence of a single monolayer of thickness 21Å which stands for the length of the 12-hydroxy stearic acid in its extended conformation (Figure 2.c). Clearly, Figures 2.a and 2.b show that the fit with such a small thickness does not reproduce the fringes experimentally observed.

Second, another possible model of arrangement at the interface is the stacking of lamellar bilayers in the plane perpendicular to the interface. This behavior has already been observed in the literature on lamellar phases of $C_{10}SO_3Na/C_{12}E_5$ mixtures [25] or on multilamellar vesicles of DDAB [26]. Here, it could occur from the unfolding of the tubes induced by the planar surface into such an 'opened' structure. This would result in a stacking of lamellar bilayers, with an interlamellar spacing close to the one observed for tubes in bulk, i.e., 270 Å (Figure 2.d). Such a layered model give rises to nice regular very large Kiessig fringes, due to the high change of neutron refractive index between the hydrogenated bilayer and $D_2O$, as observed by Salamat *et al* [25] and Mc Gillivray *et al* [26] or shown by the simulations of Figure 2.a and Figure 2.b (d = 270 Å). Though such model would correctly model the Q-position of the fringes, it is not consistent with our measurements because it completely overestimates the intensity of the fringes amplitudes. The only possibility to get such weak fringes with a lamellar stacking would come from strong fluctuations of the membranes [25]. But the fluctuations in bulk are weak in our system, as proved by SANS [6]. Moreover, the loosening of the fringes would increase with temperature, while the reverse is observed (see Figure 3a of section *III.3*).

The possibility of adsorption of multilamellar vesicles can also be dismissed because there is no transition from tubes to vesicles for 12 hydroxy stearic fatty acids in bulk solution. When such vesicles-tubes transition is observed, it is associated with the $(L_\beta)/(L_\alpha)$ transition of the chains [13]. But 12 hydroxy stearic fatty acids, multilamellar tubes are observed whether the chains are in the $L_\beta$ gel phase or in the $L_\alpha$ fluid phase [15].

We will then assume that the fringes come from the presence of multi-lamellar tubes adsorbed at the air/water interface, adsorbed on a fatty acid monolayer at the surface (see figure 2e).

The theoretical approach of the modeling of the SNR corresponding to such structure is presented in Appendix 1. It involves the same geometrical parameter as observed in bulk[15] (see Figure 2.e), i.e., an interlamellar spacing, $d$, a bilayer thickness, $e$, a tube diameter, $D$, with a number $N$ of bilayers per tube and a given surface density $\Phi_S$. N is assumed to lie between 3 and 5 from the analysis of TEM pictures and from the thickness of Bragg peaks observed in some SANS experiments of tubes in bulk [6]. The thickness, $t$, of the monolayer is fixed to 21 Å. In the model we do not consider any additional signal to take into account the remaining background coming from the off-specular scattering.

We present the SNR curve corresponding to such a calculated profile for the following experimental parameters: t = 21 Å; $d$ = 270 Å; $e$ = 41 Å, $D$ = 6000 Å; N = 2 (70%) and N=3 (30%) and $\Phi_S$ ~ 0.8. It fits rather well the data and the shape of the SNR experimental curve. However it remains below the level of the experimental SNR curve (see Figure 2.a and

Figure 2.b). We attribute this feature to the perturbative additional background arising from some remaining weak SANS scattering from the bulk solution because its influence appears as a change of baseline on the Fresnel representation of Figure 1.b, thus like an additional scattering that decays thus like $Q^{-4}$ at large Q, as obtained in SANS in the Porod regime. Thus if one adds to the calculated SNR curve a $kQ^{-4}$ term, where k is a variable constant, to roughly take into account this perturbative scattering, we recover the level of the experimental curve. Please however note that this crude additional treatment of background alters the shape of the calculated SNR curve because it has of course larger effects on the minima of the fringes than on their maxima.

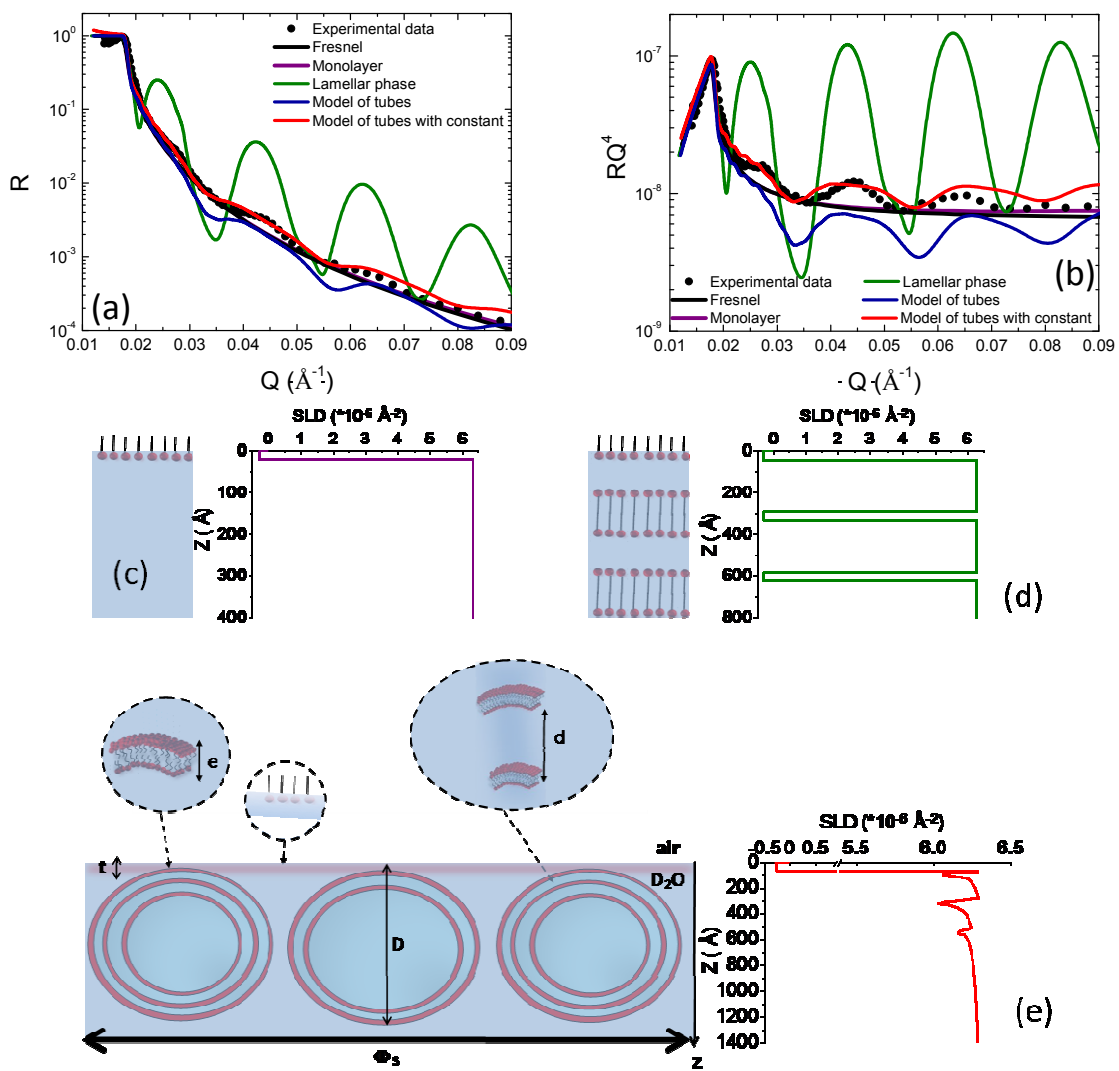

Figure 2: Comparison of the SNR results of a solution of fatty acids at 20°C with a monolayer of fatty acids, a lamellar phase and with the model of tubes adsorbed at the interface as described in the text with or without a constant in $Q^{-4}$. (a) R(Q) *versus* f(Q); (b) $Q^4$R(Q) *versus* f(Q). (c) The SLD profile used from the calculation of a fatty acid monolayer. (d) The SLD profile used from the calculation of a fatty acid lamellar phase. (e) Description of the model of tubes adsorbed at the interface with the parameters described in the text with the corresponding SLD profile.

We are conscious that our fit is performed with a large number of parameters. It is by the way easily possible to get a perfect fit of the data by a minimization of the $\chi^2$ using a free set of parameters with free variation. However, we now discuss the impact of the different parameters on the SNR curve allowing testing our model. For the sake of clarity, fits using this model obtained by varying the different parameters are shown in the supplementary information section (Figure SI.5) but we only report here the main results.

- First, the thickness of the fatty acid monolayer at the interface has been varied but does not change the overall aspect of the fit. However, we can reasonably assume it is of the order of 21 Å since it corresponds to the length of the fatty acid in its extended conformation. An accurate determination is in principle possible, even for small thicknesses, since it fixes the global envelop of the SNR curve, independently from the other parameters. However, this would imply to get a very confident fitting of the shape of the curve at large Q, at which the diffuse scattering strongly perturbs here the data.
- Second, for physical reasons, we assume that the bilayer thickness, *e*, must not be considered as a free parameter and we have thus imposed the value of 41 Å. This stands for the value measured by SANS for tubes in bulk since fatty acids are embedded in their gel phase ($L_\beta$) at 20°C.

- Third, the interlamellar spacing, *d*, is the most influent parameter because it defines the Q-position of the fringes. Then, it can be confidently determined from them and was further fixed to 270 Å.

- Then, the number of bilayers per tubes, *N*, influences the whole SNR fitting curve. The best fit is obtained for a mixing of tubes, 30% having 3 bilayers and 70%, only 2 bilayers. For N up to 3, one obtains additional oscillation superimposed to the Kiessig fringes which can be assigned to Bragg peaks coming from the interlamellar distance. For higher values of *N*, one observes additional fringes arising from the characteristic thickness of the whole tube wall, i.e., the distance between the most outer bilayer and the most inner bilayer layer of a tube.

- Also, the exact value of the tube diameter *D* has a weak importance, only its range of order has a real influence on the SNR fitting curve. As long as the tube diameter is both: (i) high enough compared to the interlamellar spacing, *d*, and (ii) low enough to consider that the bilayers are bent (*D* typically of the order of the micrometer). Indeed, for D too high, the fitting curve is that obtained for a multilamellar arrangement at the interface what has been shown to yield too intense Kiessig fringes. We took here a value of 6000 Å, as obtained from TEM pictures [6], but we have checked that values in the range 5000 Å – 8000 Å, within the errors bars of TEM, do not noticeably modify the SNR fitting curves.

- Finally, $\Phi_S$ obviously only plays on the amplitudes of the fringes. So, we have thus imposed the value of 0.8.

Thus such a modeling of the experimental data is in accordance with the presence of the multi-lamellar tubes, of same dimensions than in bulk solution, adsorbed at the air/water interface with a high packing. This high off-specular scattering we obtain is due to the presence of such tubes at the surface. SNR appears here as a powerful technique to determine their interlamellar spacing $d$ from the positions of the fringes. Indeed, even if the multi-lamellar tubes essentially contains water ($\Phi_{fc}(z')$ is at maximum 0.05 out of the surface), the jump in neutron refractive index between a bilayer of hydrogenated fatty acid and the deuterated solvent is so large that the fringes remain visible in the SNR curve.

*III.3 Evolution of the SNR data with temperature for in-cell heating*

We present in this section the evolution of the SNR experimental data with temperature. First, the solution of tubes was deposited at 20°C and heated directly within the measurement cell. The results are shown in Figure 3.a. They have been all obtained with the 1-D detector and we propose only a qualitative description based on the Q-position and on the amplitudes of the fringes.

At 40°C, the SNR curve has a very similar shape as the one obtained at 20°C. The fringes are slightly shifted towards low Q, indicating an increase of the interlamellar spacing *d*, similar to what was observed for the tubes in bulk solution [14, 15].

At 50°C, the structural behavior at the air/water interface completely changes. We recall that this temperature is the one for which, in bulk, the tube diameter increases drastically (from 0.6μm at 40°C up to 5 μm à 50°C [28]). At this temperature, the spectra recorded every 30 min evolved with time (during the first 5 hours) showing that the system has not reached equilibrium after 30 minutes (see figure 3.c). This behavior was not observed for the other temperatures.

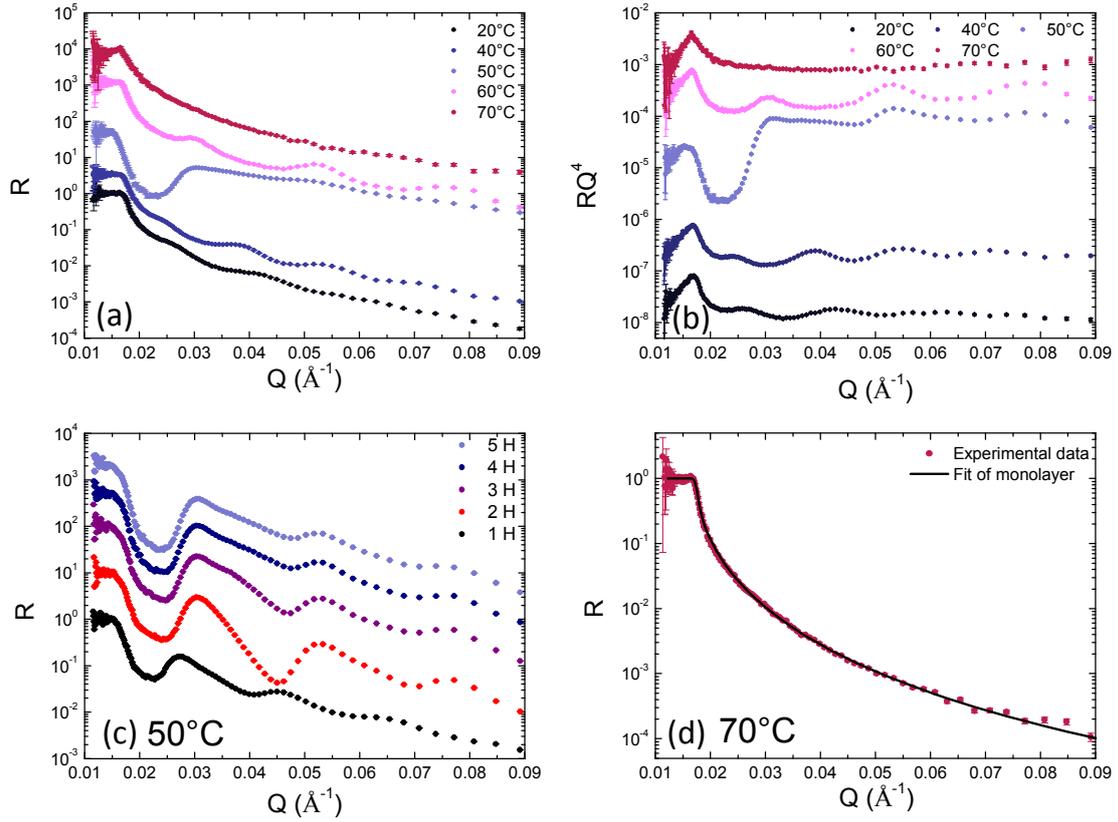

Figure 3: Evolution with temperature of the SNR curve of a solution of fatty acids at 10mg/ml in pure $D_2O$ if the solution of tubes is heated in-situ in the cell. (a) R(Q) *versus* f(Q); (b) $Q^4R(Q)$ *versus* f(Q). The data are shifted in intensity for the sake of clarity. (c) Kinetic evolution of the spectra at 50°C (each curve corresponds to a recording of one hour). (d) Fit of the experimental SNR at 70°C by a single monolayer of fatty acids.

After 1 hour, one observed strong marked fringes on the experimental curve which then resembles the SNR curve we have obtained with a model of flat stacked bilayers at the interface (see figure 2.a) and observed on lamellar systems by others [25, 26]. Here a perfect fit is obviously not possible because the system has evolved during the recording, *i.e* the SNR provides the averaged structure of the system that was not at equilibrium. However the shape prompts us to assume such stacking. We measured on that experimental curve an interlamellar spacing deduced from the position of the Kiessig Fringes and found a value (averaged thus on one hour) of around 300 Å. After 2 and 3 hours, the Kiessig fringes are even much marked (higher intensity) and were shifted toward larger wave vectors (Figure 3.c). For these two set of data, the interlamellar spacing has decreased down to around 260 Å. At longer times, the fringes go on shifting towards large Q, until they vanish progressively (Figure 3.c).

It is likely that the slow kinetic of the progressive decrease of the interlamellar spacing induces a large polydispersity on the spacing on the lamellar phases on the surface. The progressive disappearance of the fringes comes probably from the formation of surface domains with different interlamellar spacing on the surface and thus the experimental measured SNR curve, averaged on the whole surface, is a linear combination of different SNR curves with marked fringes. In the meantime, the critical wavevector $Q_c$ is shifted towards the low Q indicating that the SLD of the semi-infinite medium beneath the surface (in practice, the first thousands of Angströms) is no longer the one of the $D_2O$ solvent only. A shift towards the low Q means lowering of the SLD: there must be a huge enrichment of fatty acid materials over these first thousands of Angströms beneath surface. More quantitatively, from the position of $Q_c$ (0.015 Å$^{-1}$) we get the value of the SLD after 5 hours (4.35*10$^{-6}$ Å$^{-2}$), corresponding to a fatty acid materials fraction around 0.29 %. It is thus likely that the tubes have unfolded and coalesced to form an ordered lamellar phase stacked below the interface, oriented parallel. This finding is coherent with our previous observations in the bulk for a concentrated solution of tubes [6]. Indeed, we had shown that for a 100 mg/mL concentrated solution of tubes, the diameter increases upon heating but do not reach a maximal value as observed for a lower concentration (10 mg/mL, as used here). Rather, the diameter of the tubes indeed increases but then, the tubes 'coalesce' to form a multilamellar phase. Interestingly, upon a further heating, the tubes re-folded and their diameter decreased again. One can assume that the concentration of tubes in our present experiment at the interface is high enough to reproduce this phenomenon and that the tubes also coalesce at the interface to form flat stacked multilamellar bilayers. The decrease of the interlamellar spacing as a function of time corresponds to a dehydration between the bilayers and then, an enrichment of fatty acids at the interface. There are at least several tens of bilayers in this lamellar phase, contrary to the case of the tubes in which only 3 to 5 bilayers are involved.

At 60°C, the behavior in bulk solution is known to be close to the one observed at 40°C, i.e., the tube diameter has decreased back, probably to its initial value [28]. At the interface, one recovers a SNR curve exhibiting all the features observed at 20°C. This suggests that multilamellar tubes are again adsorbed at the interface and that the overall behavior as a function of the temperature is similar at the interface or in the bulk. The lamellar phases at the surface observed at 50°C have thus re-swollen and form again tubes adsorbed at the interface. The fringes are largely shifted towards the large Q compared to the results at 20°C and 40°C, indicating that the interlamellar spacing $d$ has decreased, exactly as observed for the tubes in bulk solution [14, 15]. This behavior as a function of the temperature supports our model in which tubes are indeed adsorbed at the interface.

Finally, at 70°C, the fringes completely vanish. At this temperature, it has been shown that tubes no longer survive in bulk solution [28], suggesting their melting into micelles. The disappearance of the fringes would thus come from the melting of the tubes at the air/water interface, letting just of single monolayer of fatty acids. To test this hypothesis, we have checked by SANS that micelles are indeed formed at 70°C. The results are presented in Supporting Information SI5. Since there are no longer tubes at the surface, the diffuse scattering has disappeared. The experimental SNR data can then be perfectly modeled by a fatty chains single monolayer, and the expected thickness of 21Å is obtained (Figure 3.d).

*III.4 Reversibility of the structural transitions with temperature*

In order to test the reversibility of the observed variations, the temperature has been decreased from 60°C down to 40°C, and then heated again up to 70°C, with one measurement every 10°C. Results are presented in Figure 4. For a given temperature, there are only slight differences between the results corresponding to the first increase of temperature, the decrease of temperature and the second increase of temperature. All structural transitions at the air/water interface are then quasi-completely reversible! It is not easy to attribute the small differences between the curves to a real change in structural behavior, rather than structural changes induced by the experimental protocol chosen. In particular, the sample has been left for more than 70 hours within the cell (6 hours per measurement at a given temperature and 30 minutes for a temperature change and its equilibrium), mainly at rather high temperature, which results in the formation of a lot of $D_2O$ vapor within the cell. As $D_2O$ is a strong neutron scatterer, this vapor may alter a bit the quality of the data.

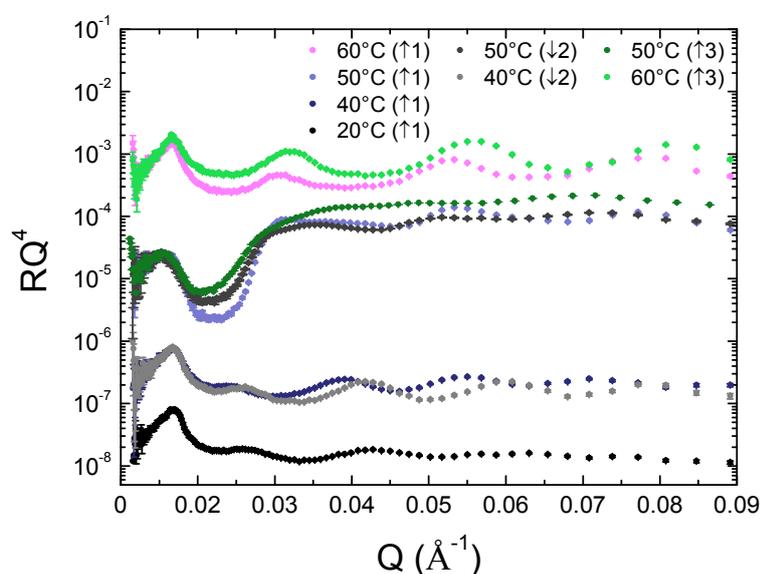

Figure 4: Comparison between the evolutions of the SNR curves of a solution of fatty acids at 10mg/ml in pure $D_2O$ if the solution of tubes is heated directly in-situ in the cell measurement during two successive increases of the temperature ((1) first increase, (2) decrease and (3) second increase). The data are shifted in intensity for the sake of clarity.

*III.5 Influence of the thermal history of the sample*

The results of the previous section have shown that the structural behavior of the tubes at the air/water interface is very similar to the one in bulk solution, except the specific behavior at 50°C. A series of experiment was done with the sample thermalized ex-situ out of the measurement cell at the desired temperature. Then, this heated sample has been deposited in the pre-heated measurement cell allowing testing the influence of the thermal history of the sample on the arrangements of fatty acids at the surface. The results are gathered figure 5 on which are superimposed the data of the previous section.

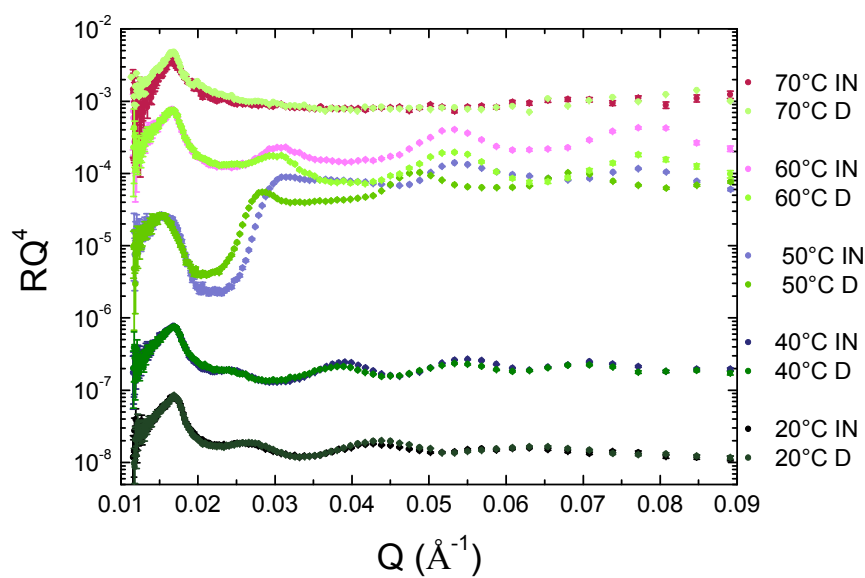

Figure 5: Evolution of the SNR curve of a solution of fatty acids at 10mg/ml in pure $D_2O$ if the solution of tubes is heated directly in-situ in the cell measurement during experiments with temperature (IN) or if the solution of tubes is heated ex-situ out of the measurement cell at the desired temperature, prior to the deposit in the pre-heated measurement cell (D). The data are shifted in intensity for the sake of clarity.

Whatever the temperature, the results from the in-situ heating protocol and from the deposit at desired temperature protocol are identical! This definitely proves that the rearrangements of the structure of the tubes at the air/water interface are completely driven by the structural behavior of the tubes in bulk, except for the specific behavior at the interface at 50°C.

III.6 Surface behavior versus bulk behavior

All the results from SNR unambiguously prove that the multi-lamellar tubes of fatty acid chains are present at the air/water interface, whereas one could have expected an 'unfolding' of the tubes induced by the hydrophobic planar air/water interface. The lack of history-dependence of the structures at the air/water interface, and the fact that the structure in bulk and at the air/water interface follows the same trends, suggest that the surface structure is driven by the bulk structure.

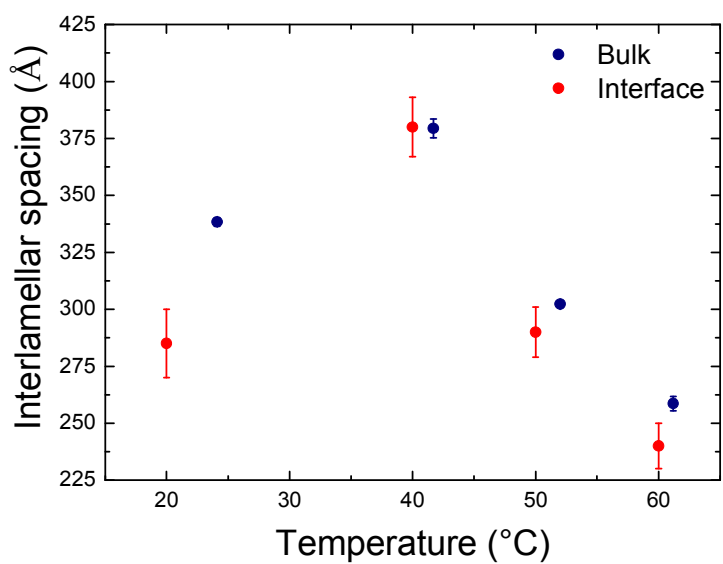

Figure 6: Comparison between the evolution with temperature of the interlamellar spacing of tubes either in bulk or at the air/water interface.

Remarkably, the interlamellar spacing $d$ has exactly the same values, within the experimental errors, in bulk solution[6, 14, 15] and at the air/water interface, as shown by Figure 6. The values of $d$ reported for the air/water interface have been obtained from simulations of pure lamellar phases at the interface, which gives exactly the same interlamellar spacing as in tubes when their minima match the ones of the fringes (see Figure 2.b), as most experiments have been performed with the 1-D detector that do not allow an unambiguous modeling with tubes. The value at 50°C corresponds to the one of the planar stacked lamellar phase measured during the first hour of measurement for such a temperature. The bulk values of *d* are obtained from the positions of the Bragg peaks of the SANS experiments[28,6, 14, 15].

The interlamellar spacing follows the same trends in bulk and at the air/water interface, confirming from a quantitative point of view the narrow link between surface and bulk.

**Conclusion**

We have shown here by specular neutron reflectivity that multilamellar tubes of ethanolamine salt of the 12-hydroxy stearic acid adsorb at the air/water interface. Similarly to what observed in bulk solution, the tubes have a temperature tunable diameter at the interface. Moreover, they have almost the same temperature dependence than in bulk. We have pointed a peculiar behavior around 50°C at which the tubes unfold and coalesce to form a multilamellar phase at the interface. It is also remarkable that this set of structural transitions at the interface is also quasi-completely reversible at surface as in bulk, whether the temperature is increased or decreased. The complete structural behavior of the tubes at the interface and in bulk solution is summarized in scheme 1. Finally, let us remark that the tuning of the structure of the multilamellar tubes at the interface by the temperature provides to the system exceptional potential for its interfacial properties (foams and emulsions).

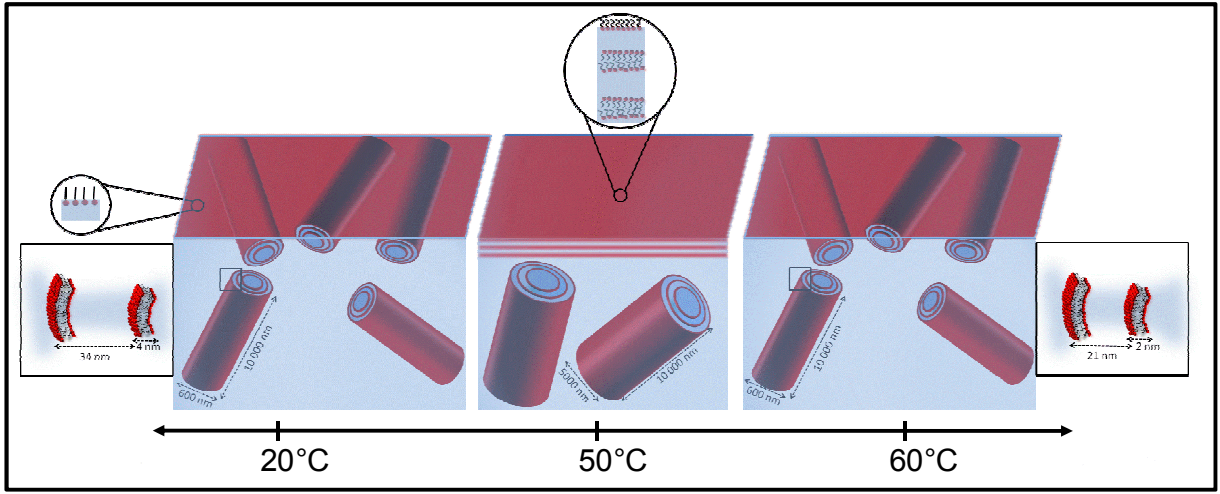

Scheme 1: Structural behavior of the tubes in bulk and at the air/water interface as a function of the temperature.

## Appendix 1 Theoretical Scattering Length Density profile of multilamellar tubes adsorbed on a planar interface.

We define the zero position of the z-axis perpendicular to the surface at the ordinate corresponding to the contact between the monolayer and the tubes. $\Phi_S$ is defined as the surface fraction occupied by tubes of diameter D at z = D, *i.e* $\Phi_S$ is 1 if all tubes are in contact. The volume fraction of fatty chains only due coming to the sole tubes $\Phi_{fc\_tubes}(z)$ writes:

$$\Phi_{fc\_tubes}(z) = \Phi_S \sum_{n=0}^{n=N-1}\left[A_n(z) - B_n(z)\right] \qquad (1)$$

where $A_n(z)$ and $B_n(z)$ correspond respectively to the outer position and inner position of the $n^{th}$ bilayer in the tubes, starting from the surface.

We have thus $A_n(z) = 0$ for $z < nd$ ;

$$A_n(z) = \left( \left( \frac{D}{2} - nd \right)^2 - \left| z - \left( \frac{D}{2} \right) + nd \right|^2 \right)^{1/2} \text{ for } nd < z < D\text{-}nd;$$

$A_n(z) = 0$ for $z > D - nd$;

and $B_n(z) = 0$ for $z <$ nd + $e$;

$$B_n(z) = \left( \left( \frac{D}{2} - nd - e \right)^2 - \left| z - \left( \frac{D}{2} \right) + nd + e \right|^2 \right)^{1/2} \text{ for } nd + e < z < D\text{-}nd - e;$$

$B_n(z) = 0$ for $z > D$ -n$d$ - $e$;

Thus if we define $z'$ by $z$ – t ($z'$ = 0 at the air/monolayer), the volume fraction of fatty chains $\Phi_{fc}(z')$ in the plane perpendicular to the interface becomes :

$\Phi_{fc}(z') = 1$ for $z' < t$;

$\Phi_{fc}(z') = \Phi_{fc\_tubes}(z)$ for $t < z' < D + t$;

$\Phi_{fc}(z') = 0 \ z' > D + t$; (2)

And the SLD profile becomes $N_b(z')$ :

$N_b(z') = N_{b\_fc}\Phi_{fc}(z') + N_{b\_solv}(1 - \Phi_{fc}(z'))$ (3)

where $N_{b\_fc}$ and $N_{b\_solv}$ are respectively the scattering length density of the fatty acid and of the solvent.

**Aknowledgements:** We thank Alain Menelle (CEA, Laboratoire Léon Brillouin, France) for stimulating discussions on the NR experiments. We gratefully acknowledge Laurence Navailles and Frédéric Nallet (CNRS, Centre de Recherche Paul Pascal) for fruitful discussions on fatty acids tubes.

For Table of Contents Use Only

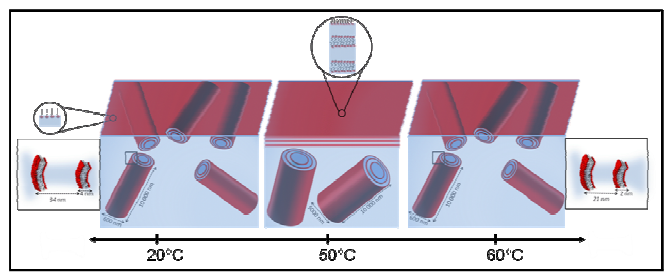

Adsorption of multilamellar tubes with a temperature tunable diameter at the air/water interface

*Anne-Laure Fameau[1,2], Jean-Paul Douliez[2], François Boué[1], Frédéric Ott[1] and Fabrice Cousin[1,*]*

# Supporting Information

Adsorption of multilamellar tubes with a temperature tunable diameter at the air/water interface: a process driven by the bulk properties

A-L.Fameau[1,2], J-P. Douliez[2], F. Boué[1], F.Ott[1], B. Novales[2], and F. Cousin[1,*]

[1] Laboratoire Léon Brillouin, CEA Saclay, 91191 Gif sur Yvette Cedex, France
[2] UR1268, INRA Nantes, Biopolymères Interactions Assemblages, rue de la Géraudière, 44316 Nantes, France

(*) To whom correspondence should be addressed: Fabrice.Cousin@cea.fr

1) <u>**3 different areas of integration of the specular signal**</u>

We show in Figure 1, the 3 different areas of integration of the specular signal with the 2-detector corresponding to the reflectivity curves shown in the paper. We clearly observe a very intense diffuse scattering. Thus, if the integration area is too large on the 2-detector, we integrate some of the off-specular signal, interfering with the exact determination of the specular scattering.

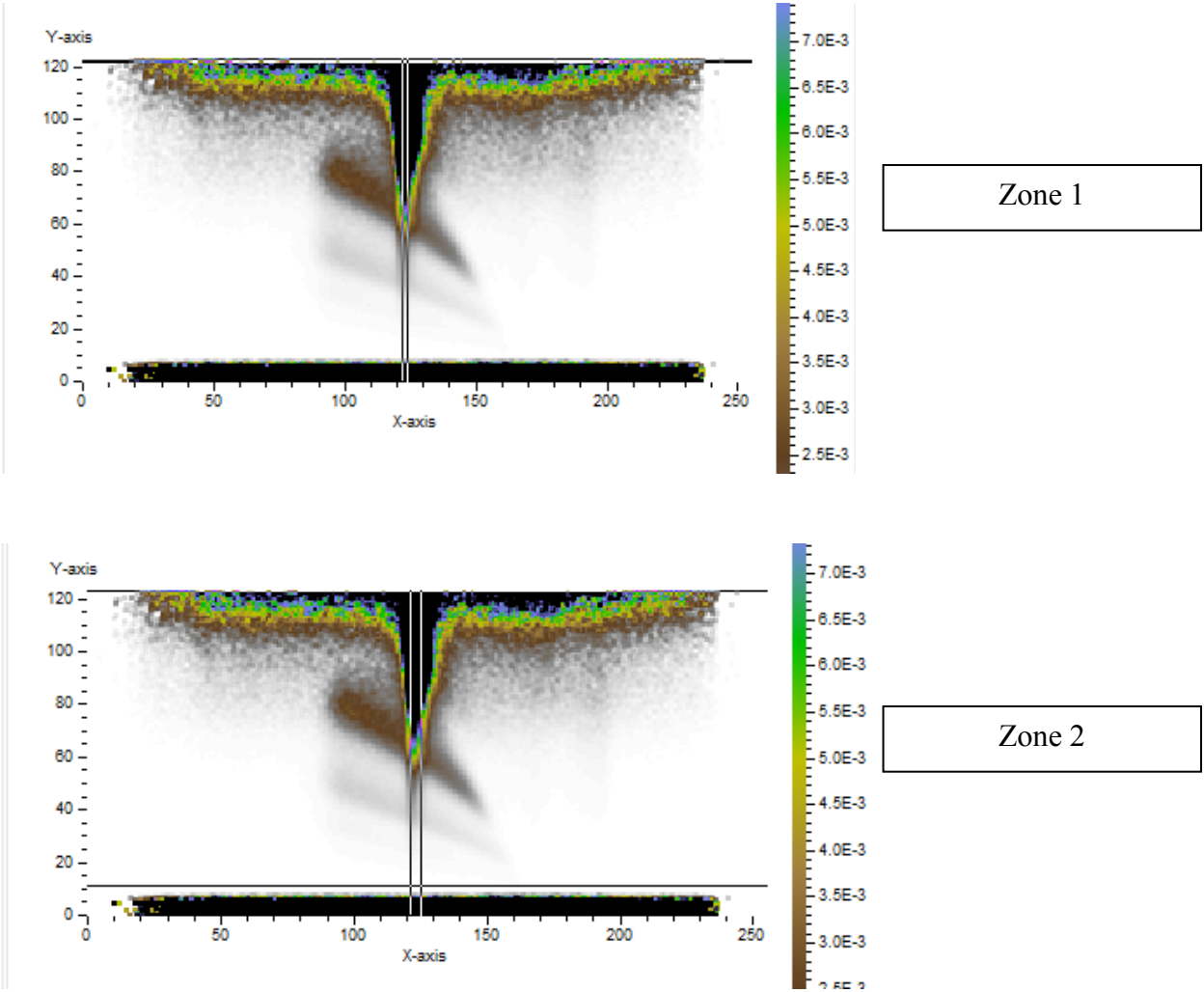

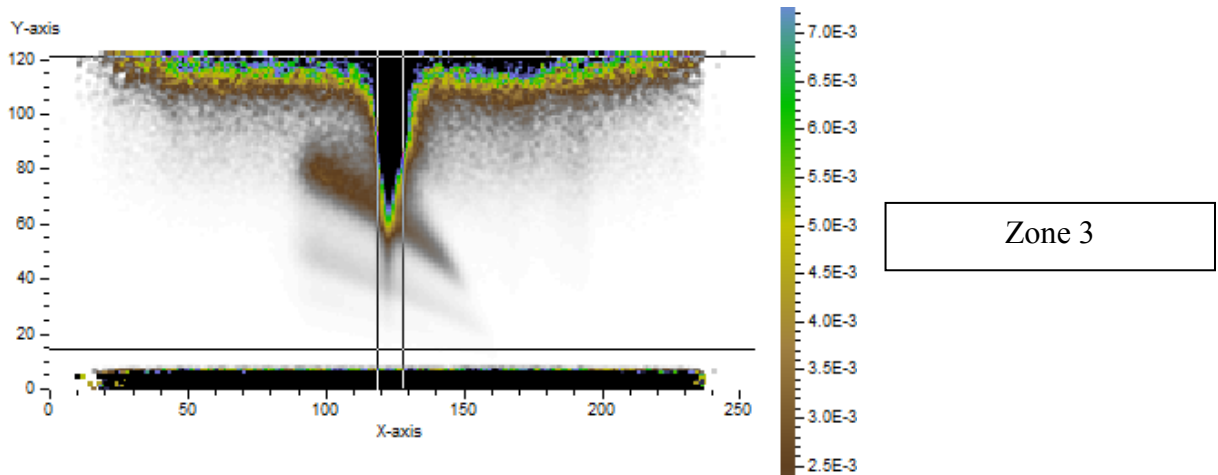

Figure 1: Comparison between the 3 different areas of integration of the specular signal with the 2-detector.

**2) Effect of various parameters on the model of tubes adsorption: the number of bilayer by tube N (a), the interlayer distance d (b), the given surface density of tubes $\Phi_{S\_tubes}$, (c), the constant (d)**

In this paper, we model the experimental SNR curve with tubes adsorbed on the monolayer at the interface with same geometrical parameter as observed in bulk. We show here the impact of the different parameters on the SNR curve to demonstrate that we can get a real confidence in our physical description of the data as a model of tubes adsorbed on the surface. We realized different calculations of tubes adsorbed at the interface where each parameters is varied independently from the others in order to show the influence of a variation of each parameter on the resulting SNR curve. In the figure 2a., we show the effect of the number of bilayer by tubes N. Clearly, it appears that from N=3, we have no longer tiny Kiessig fringes. In the figure 2b, we show the effect of the interlamellar spacing *d*. This parameter fixes unambiguously the q-position of the minima fringes. Then, we studied the effect of the given surface density of tubes $\Phi_{S\_tubes}$. As you can see on the figure 2c, this parameter only plays a role on the amplitude of the fringes. In the figure 2d, we show the model of tube with or without a constant. In fact, the model of tubes matches very well the shape of the experimental SNR curve but it remains below the level of the experimental SNR curve because of the perturbative background arising from the diffuse scattering. The intensity of such scattering as a function of Q follows the same trends as the specular intensity and must decay thus like $Q^{-4}$ following the Fresnel decay. Thus if one adds to the calculated SNR curve a $kQ^{-4}$ term, where k is a variable constant to roughly take into account the diffuse scattering, one recovers the level of SNR of the experimental curve. However, we can see that this rude additional treatment of background alters the shape of the calculated SNR curve because it has a more important weight for the minima of the fringes than for the maxima.

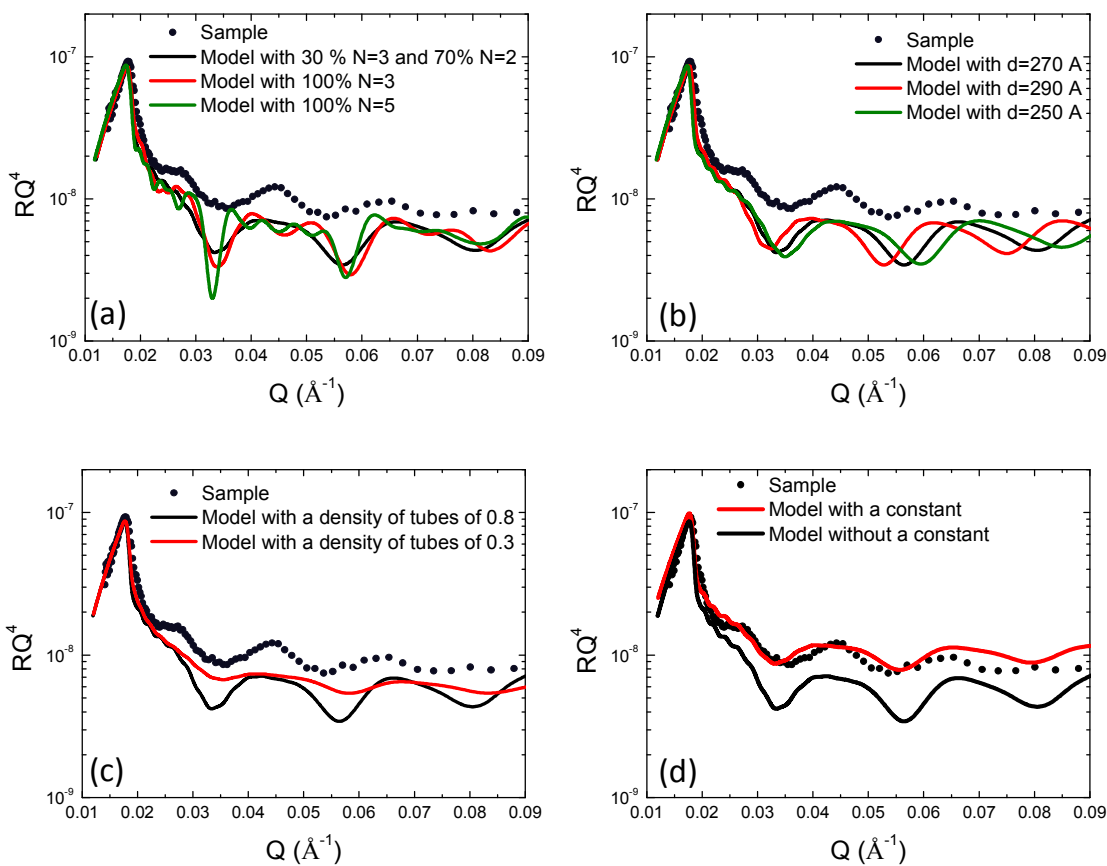

Figure 2: Comparison of the SNR results of a solution of fatty acids at 20°C with a monolayer of fatty acids, with the model of tubes adsorbed at the interface as describes in the paper by varying separately each parameters. (a) Effect of the number of bilayer by tubes N. (b) Effect

of the value of the interlayer spacing. (c) effect of the given surface density of tubes $\Phi_{S\_tubes}$.

(d) Effect of the constant.